\documentclass[fleqn,12pt,twoside]{article}
\usepackage{espcrc1,epsfig}

\usepackage{graphicx}
\usepackage[figuresright]{rotating}

\newcommand{\AmS}{{\protect\the\textfont2
  A\kern-.1667em\lower.5ex\hbox{M}\kern-.125emS}}

\title{Thermal phase transition in the NJL model of QCD}

\author{C.G. Strouthos\address[MCSD]{Division of Science and Engineering, \\ 
         Frederick Institute of Technology, Nicosia 1303, Cyprus.}} 
       
\begin{document}
\maketitle

\begin{abstract}
We present results from numerical studies of the NJL model with two massless quarks at 
nonzero temperature. We show that the model undergoes a second order chiral phase transition 
which belongs to the $3d$ $O(4)$ spin model universality class. 
\end{abstract}

\section{INTRODUCTION}

Massless QCD with two flavors undergoes a chiral symmetry restoration
transition which is believed to be of second order. According to arguments
by Pisarski and Wilczek \cite{pw} the universality class of the transition is
that of the dimensionally reduced $3d$ $O(4)$ Heisenberg spin model.
The reasoning behind their conjecture is that quantum fluctuations become
irrelevant near the critical temperature $T_c$. As $T \rightarrow T_c$ the correlation length $\xi$ diverges 
and the characteristic mode is $\omega_c \sim 0$. The fermions are expected to decouple from the scalar sector 
because they do not have zero modes. The energy of the  zero point quantum fluctuations of characteristic
frequency $\omega_c$,  $\hbar \omega_c$ is much less than the thermal energy per degree
of freedom $k_BT$. Therefore, in this regime one observes the critical behaviour of a classical
(thermal) transition.
However, there is a possible loophole in this scenario. In fermionic models the mesons are composite
$f\bar{f}$ states whose density and size increase as $T \rightarrow T_c$.
This may imply that the fermionic sub-structure is apparent on physical length scales and as a 
consequence we may have a maximal violation of the bosonic character of mesons near $T_c$.
If this is the case then the quarks become essential degrees of freedom irrespective of how heavy
they are.

Since the problem of chiral symmetry breaking and its restoration is intrinsically non-perturbative
the number of techniques available is limited and most of our knowledge about this phenomenon comes from lattice
simulations. Various lattice results indicate that the QCD chiral phase transition at $T \neq 0$ is second 
order \cite{order}. However, due to problems with finite size effects near the critical temperature and the proper 
inclusion of fermions on the lattice, these calculations are still not fully satisfactory. Therefore, an important role
is played by model field theories which incorporate various basic features of QCD and they are much easier to deal with.
Such a theory is the Nambu$-$Jona-Lasinio (NJL) model. The NJL model was introduced in the sixties as a theory 
of interacting nucleons \cite{nambu} and later it was reformulated in terms of quark degrees of freedom.
In the continuum, the model is described by the Euclidean Lagrangian density
\begin{equation}
{\cal L}  = \overline{\psi}_i \partial\hskip -.5em / \psi_i - \frac{g^2}{2N_f}
 \left[\left(\overline{\psi}_i\psi_i\right)^2-\left(\overline{\psi}_i
\gamma_5 \vec\tau\psi_i\right)^2\right],
\label{Lcont}
\end{equation}
where $\vec{\tau}\equiv(\tau_1,\tau_2,\tau_3)$ are the Pauli spin
matrices which run over the internal $SU(2)$ isospin symmetry.
The index $i$ runs over $N_c$ fermion colours and $g^2$ is the coupling
constant of the four-fermion interaction.
The model is chirally
symmetric under $SU(2)_L\otimes SU(2)_R$;
$\psi\to (P_LU+P_RV)\psi$,
where $U$ and $V$ are independent global $SU(2)$ rotations and the
operators
$P_{L,R}\equiv\frac{1}{2}(1\pm\gamma_5)$
project onto left
and right handed spinors respectively. It is also
invariant under $U(1)_V$
corresponding to a conserved baryon number.
The theory becomes easier to treat, both analytically and numerically, if we
introduce scalar and pseudo-scalar
fields denoted by $\sigma$ and $\vec\pi$ respectively.
The bosonised Lagrangian is
\begin{eqnarray}
{\cal L}  =  \overline{\psi}_i\left(\partial\hskip -.5em / +\sigma+i\gamma_5
\vec\pi \cdot \vec\tau\right)\psi_i
+\frac{N_f}{2g^2}\left(\sigma^2+\vec\pi \cdot \vec\pi\right). 
\end{eqnarray}
The dynamic generation of fermion masses brought about by the breaking of chiral symmetry 
at $g^2$ larger than a critical $g_c^2$ is one of the important properties of the model.
This implies the creation of the Nambu-Goldstone bosons ($\pi$) and the creation of the chiral partner
sigma ($\sigma$) which are $f\bar{f}$ bound states. 
The interaction strength has a mass dimension -2, implying that the model is non-renormalizable. It works well
in the intermediate scale region as an effective theory of QCD with a momentum cutoff comparable 
to the scale of chiral symmetry breaking $\Lambda_{\chi SB} \simeq$ 1GeV.
Various authors demonstarted the success of the model as a quantitative theory of hadrons (for a review see e.g. 
\cite{hatsuda}). 
It was also shown that the model has an interesting phase diagram in the 
(temperature, quark chemical potential $\mu$) plane, in close agreement with predictions from other effective
field theories of QCD. 
In the Hartree-Fock approximation, the theory has a tricritical point in the $(T,\mu)$ plane \cite{tricr} 
when the quark current mass, the coupling $g$ and the momentum cut-off $\Lambda$ 
are chosen to have physically meaningful values.
Recently, lattice simulations showed that the ground-state at high $\mu$ 
and low $T$ is that of a traditional BCS superfluid \cite{hands2004}.
It was also shown numerically that the $d$-dimensional $Z_2$ symmetric model undergoes a finite temperature phase 
transition which belongs to the $(d-1)$-dimensional Ising model \cite{strouthos,chandra} 
implying that the dimensional reduction 
scenario is the valid description of the thermal transition. Here we study the case where the chiral symmetry is 
continuum. 

\section{RESULTS}
We performed lattice simulations of the NJL model with the Hybrid Monte Carlo
method. Details about the staggered fermion action used in our simulations can be found in \cite{hands2004}.
The asymmetric lattice has $L_s$ lattice spacings $a$ in spatial directions, $L_t \ll L_s $ lattice spacings in the 
temporal direction and volume $V \equiv L_s^3 L_t$. The temperature is given by $T=1/(L_ta)$.
On a finite volume and with the quark bare mass set to zero,
the direction of symmetry breaking changes over the course of the run so the chiral condensate averages to zero 
over the ensemble. Another option is to introduce an effective order parameter 
$\Phi \equiv \frac{1}{V}\sqrt{(\sum_x \sigma(x))^2 + \sum_i (\sum_x \pi_i(x))^2}$, which in the thermodynamic limit 
is equal to the true order parameter
$\Sigma = \langle \sigma \rangle$ extrapolated to zero quark mass.
In order to study the critical behaviour on lattices available to us we use the finite size scaling (FSS)
method, which is a well-established tool for studying critical properties of phase transitions. 
On a finite lattice the correlation lenght $\xi$ is limited by the size of the system and consequently
no true criticality can be observed. 
The dependence of a given thermodynamic observable, $P$, on the
size of the box is singular and according to the FSS
hypothesis is given by:
\begin{equation}\label{fssX}
P(t,L_s) = L_s^{\rho_P/\nu}Q_P(tL_s^{1/\nu}),
\end{equation}
where $t \equiv (\beta_c-\beta)/\beta_c$,
$\nu$ is the standard exponent of the correlation length
and $Q_P$ is a scaling function, which is not
singular at zero argument.
The exponent $\rho_P$ is the standard critical exponent
for the quantity $P$. 
Studying the dependence of $P$ on $L_s$
and using eq.~(\ref{fssX}) one can determine such exponents.
We performed simulations with $N_c=8$, fixed $L_t=4$ and $L_s=8,...,54$ at three or four values of $\beta$ 
for each lattice size and then used
histogram reweighting techniques to calculate the values of the observables at nearby couplings.
The temperature $T$ is varied by varying the lattice spacing $a(g)$.
Near $\beta_c$ the expansion of the FSS scaling relation for the susceptibility 
$\chi=V\langle \Phi^2 \rangle$ is:
\begin{equation}\label{susc}
\chi = (a_0 + a_1t + a2t^2 + ...) + (b_0 + b_1tL_s^{1/\nu} + b_2t^2L_s^{2/\nu} +...)L^{\gamma/\nu}.
\end{equation}
For large enough $L_s$ one can neglect the terms in the first parenthesis. 
By fitting our data to eq.~(\ref{susc})
we get: $\beta_c=0.52579(3)$, $\nu=0.74(3)$, $\gamma/\nu=1.95(1)$. These results are in agreement with the 
$3d$ $O(4)$ spin model exponents \cite{kanaya} $\nu=0.748(9)$ and
$\gamma/\nu=1.9746(38)$. 

A method usually used to determine $\beta_c$ is by measuring the
Binder cumulant,
\begin{equation}
U_L \equiv 1 - \frac{1}{3} \frac{\langle \Phi^4 \rangle}{\langle \Phi^2 \rangle^2} .
\end{equation}
As a consequence of eq.~(\ref{fssX}), near $\beta_c$,
$U_L=f(tL^{1/\nu})$. In the large $L_s$ limit $U_L$ has a universal value $U_*$ at $\beta_c$.
Close to $\beta_c$,  
$U(L_s,\beta) \simeq U_* + a_1(\beta_c-\beta)L^{1/\nu} + a_2(\beta_c-\beta)^2L^{2/\nu}$.
The value of $U_*$ in the classical $3d$ $O(4)$ universality is 0.636(1).
Fitting the above relation to data generated on lattices with $L_s=30,40,54$ we get  $\nu=0.69(4)$, $U_*=0.616(2)$ and 
$\beta_c=0.5261(1)$. The measured values of $\nu$ and $U_*$ are close to their respective values in 
the $3d$ $O(4)$ spin model.

It has been shown analytically that in the infinite $N_c$ limit the theory undergoes a second order phase
transition with Landau-Ginzburg mean field (MF) exponents for any dimensionality $d$ \cite{kocic}. 
The reasoning behind this behaviour is that bosonic fluctuations are neglected, whereas the fermions
decouple because their non-zero Matsubara frequency acts as a mass term of order $\pi T$ for the
$(d-1)$-dimensional components of the field. 
However, as we showed in the previous paragraph for $N_c=8$ the dimensional reduction scenario is the correct
description of the transition. 
It has been shown in \cite{strouthos} that although the fermions do not affect the actual critical 
behaviour they do affect the way the universal behaviour sets in. Close to $T_c$, MF theory is expected to 
break down and 
in the case of $d=4$ the non-trivial scaling region is suppressed by a factor $\sim N_c^{-1}$ \cite{strouthos}.
\begin{figure}[t!]
\begin{minipage}[t]{75mm}
\epsfig{file=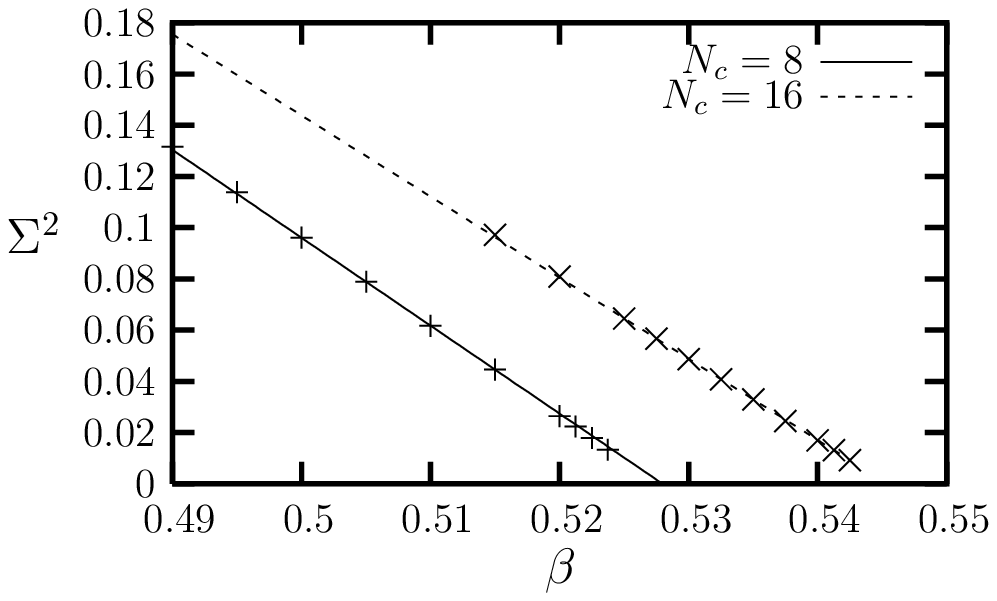, width =7.5cm}
\caption{$\Sigma^2$ vs $\beta$ for $N_c=8,16$.}
\label{fig:fig1}
\end{minipage}
\hspace{\fill}
\begin{minipage}[t]{75mm}
\epsfig{file=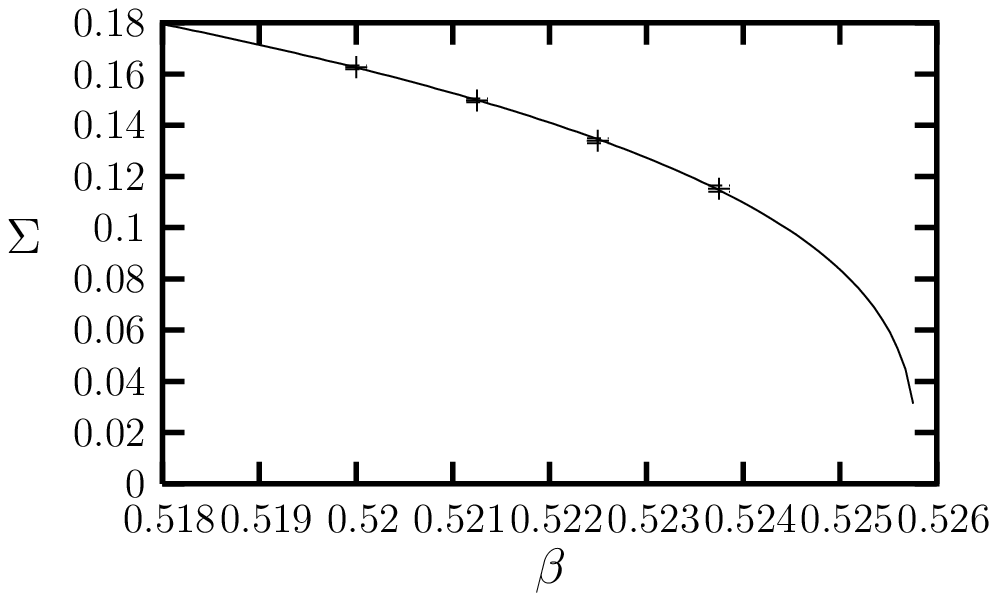, width =7.5cm}
\caption{$\Sigma$ vs $\beta$ close to $\beta_c$ for $N_c=8$.}
\label{fig:fig2}
\end{minipage}
\end{figure}

We perfomed simulations in the broken phase on $4 \times 54^3$ lattices with $N_c=8$ and $N_c=16$.
In the MF region the order parameter $\Sigma$ is equal to the fermion mass and therefore acts
as an inverse correlation length. In Fig.~\ref{fig:fig1} we plot $\Sigma^2$ versus $\beta$. The data in the MF region 
fits well to straight lines because in this region the critical exponent for $\Sigma$ is $\beta_{\rm mag}=1/2$.
For $N_c=8$ the quality of the linear fit deteriorates significantly when we include the $\beta=0.520$ point. 
For $N_c=16$ all the $\beta > 0.515$ data give a high quality fit, implying that
in this case the crossover from the MF region to the $3d$ $O(4)$ region   
occurs at larger correlation lengths than for $N_c=8$. This result verifies that 
the $3d$ $O(4)$ scaling region is suppressed at large $N_c$. We also fitted the $N_c=8$
data for $\beta =0.520 - 0.5225$ (see Fig.~\ref{fig:fig2}) to the scaling relation 
$\Sigma = {\rm const.}(\beta_c - \beta)^{\beta_{\rm mag}}$
with fixed $\beta_c=0.5260$ and extracted $\beta_{\rm mag} = 0.360(15)$ which is close to the $3d$ $O(4)$ exponent
$\beta_{\rm mag}=0.3836(46)$ \cite{kanaya}. 
 
\section{SUMMARY}   
We showed numerically that the NJL model of QCD with two massless quarks undergoes a second order 
chiral phase transition which belongs to the $3d$ $O(4)$ spin model universality. Furthemore, as shown 
some time ago in the $(2+1)d$ $Z_2$-symmetric Gross-Neveu model \cite{strouthos} the actual critical region 
is suppressed if the number of fermion species in the system is large. It will be interesting to check with comparable
precision whether this scenario is the correct description of the QCD thermal phase transition.
\\
\\
Discussions with S. Chandrasekharan, S. Hands and J. Kogut are greatly appreciated.

\end{document}